\documentclass{aa}
\usepackage[dvips]{graphicx}
\newcommand{\crat}{$^{12}$C/$^{13}$C~}
\newcommand{\msun}{$M_\odot$}
\newcommand{\etal}{et al.~}
\newcommand{\simless}{\mathbin{\lower 3pt\hbox
 {$\rlap{\raise 5pt\hbox{$\char'074$}}\mathchar"7218$}}} 
\newcommand{\simgreat}{\mathbin{\lower 3pt\hbox
 {$\rlap{\raise 5pt\hbox{$\char'076$}}\mathchar"7218$}}} 

\begin{document}

\thesaurus{09.16.1; 13.03.19.3; 10.05.1}
\title{Measurements of \crat in planetary nebulae: 
implications on stellar and Galactic chemical evolution}
\author{F. Palla\inst{1}, R. Bachiller\inst{2}, L. Stanghellini\inst{3,4}\thanks{Affiliated with the Astrophysics Division, Space Science 
Department of ESA}, M. Tosi\inst{3}, D. Galli\inst{1}} 
\institute{Osservatorio Astrofisico di Arcetri, Largo E. Fermi 5, 
   I-50125 Firenze, Italy \and
	   Observatorio Astron\'omico Nacional (IGN), Campus Universitario, 
   Apdo. 1143, E-2880 Alcal\'a de Henares (Madrid), Spain \and
	   Osservatorio Astronomico di Bologna, Viale Berti Pichat 6/2, 
   I-40126 Bologna, Italy \and
	   Space Telescope Science Institute, 3700 San Martin Drive, 
  Baltimore MD 21218, USA }
\offprints{F.~Palla}
\mail{palla@arcetri.astro.it}
\authorrunning{Palla et al.}
\titlerunning{Measurements of the \crat in PNe}
\date{Received date: 16 April 1999; accepted date: 8 November 1999}
\maketitle
\begin{abstract}

We present the results of a study aimed at determining the
\crat ratio in two samples of planetary nebulae (PNe) by
means of millimeter wave observations of $^{12}$CO and $^{13}$CO. The first
group includes six PNe which have been observed in the $^3$He$^+$
hyperfine transition by Balser et al. (1997); the other 
group consists of 22 nebulae with rich
molecular envelopes.  We have determined the carbon isotopic ratio in 14
objects, 9 of which are new detections. The results indicate a range 
of values of $^{12}$C/$^{13}$C between 9 and 23.
We estimate the mass of the progenitors of the PNe of our sample and
combine this information with the derived \crat isotopic ratios to test
the predictions of stellar nucleosynthesis models.

We find that the majority of PNe have isotopic ratios 
below the values expected from current standard asymptotic giant branch
models in the mass range of interest.  We suggest that the progenitors of the 
PNe must have undergone a non-standard mixing process during their red giant 
phase and/or asymptotic giant phase, 
resulting in a significant enhancement of the $^{13}$C
abundance in the surface layers. Our study confirms a similar behaviour
inferred from spectroscopic observations of field population II stars
and globular cluster giants, and extends it to the final stages of
stellar evolution.  Finally, we discuss the implications of our results
on models of Galactic chemical evolution of $^3$He and $^{12}$C/$^{13}$C.

\end{abstract}

\keywords{planetary nebulae: general -- radio lines: ISM -- Galaxy: evolution}

\section{Introduction}

In the PN phase, stars more massive than 0.8~\msun\ return to the ISM
material that has been processed in the stellar interior. This matter
mixes with the surrounding medium and modifies the original abundances
of elements. The contribution of PNe to the Galactic chemical evolution
is particularly important for $^3$He which, together with deuterium,
plays a fundamental role in testing the standard Big Bang
nucleosynthesis model. While the evolution of deuterium is well
understood, that of $^3$He still encounters serious problems which cast
doubts on the usefulness of this isotope as a test of Big Bang
nucleosynthesis models (e.g. Galli et al.~1995; for a different
opinion on deuterium see Mullan \& Linsky~1999). Observations of $^3$He
toward PNe and H{\sc ii} regions have resulted in abundances that
differ by almost two orders of magnitude:  $^3$He/H~$\sim 10^{-3}$ in
PNe (Rood et al.~1992, Balser et al.~1997), and
$^3$He/H~$\sim 10^{-5}$ in H{\sc ii} regions (Balser et al.~1994, Rood
et al.~1995). The latter value is also representative of the $^3$He
abundance in the presolar material (Geiss~1993) and the local ISM
(Gloeckler \& Geiss~1996).  However, the abundance in PNe agrees with
the predictions of standard stellar evolution models for stars of mass
1--1.5~\msun (see the review by Rood et al.~1998).  The main question
is then: if low-mass stars are net producers of $^3$He and return it to
the ISM during the PN phase, why don't we observe a much higher
abundance in H{\sc ii} regions and in the solar system material, as
all standard Galactic evolutionary models predict (see e.g. Tosi~1996)?

An interesting solution to this problem involves the existence of a
nonstandard mixing mechanism (or Cool Bottom Processing, hereafter CBP) which
operates during the red giant phase of stars with $M\simless 2$~\msun.
In addition to decreasing the amount of $^3$He in the stellar envelope,
this process affects the abundances of other important elements,
including carbon, as first suggested by Hogan~(1995), Charbonnel~(1995)
and Wasserburg et al.~(1995).  In particular, the ratio
of \crat in the envelope is predicted to be much {\it lower} than in
the standard case. For a 1~\msun\ star, the predicted ratio is about 5
against the standard value of 25--30 in the red giant branch (RGB) and
of 20--40 in the asymptotic giant branch (AGB). However, the
discrepancy becomes larger for more massive stars where the \crat
ratio can reach $\sim$100 in the AGB phase
(Charbonnel~1995;
Weiss et al.~1996; Forestini \& Charbonnel~1997,
hereafter FC; van den Hoek \& Groenewegen~1997, herafter HG;
Marigo~1998; Boothroyd \& Sackmann~1999, hereafter BS).

From an observational viewpoint, it is important to obtain accurate
measurements of the isotopic ratio in those PNe where the $^3$He
abundance has been determined. Should these objects show a high value of
$^{12}$C/$^{13}$C, then no modifications to the standard stellar models
would be required. Otherwise, one has to invoke another selective
process (mixing, diffusion etc.) that operates on some isotopes but not
on $^3$He. However, the number of PNe with $^3$He measurements is small
(e.g. Balser et al. 1999), whereas the suggested physical processes
should be quite general and should affect the nucleosynthetic yields of
all stars of mass less than $\sim 2$~\msun.  Hence, it is critical to
measure the carbon isotopic ratio in a sample of PNe as large as
possible.

The molecular envelopes of PNe have been studied extensively at near
infrared and millimeter wavelengths (see e.g. Kastner et al.~1996,
Huggins et al.~1996, Bachiller et al.~1997). These observations have
shown that massive envelopes ($\simgreat 10^{-2}$~\msun) containing a
rich variety of molecular species are commonly found around PNe. CO is
the most widely observed species, and the \crat isotopic ratio has been
measured toward several PNe (Bachiller et al.~1989, 1997; Cox et
al.~1992). These initial studies have shown that the \crat ratio is in
the range 10--20.

Our project consists of two parts. In the first one, we have carried
out high quality observations of $^{12}$CO and $^{13}$CO in six PNe
that have been searched for $^3$He emission.  In the second run, a
larger sample of nebulae with strong $^{12}$CO line emission has been
observed in the $^{13}$CO lines in order to determine the isotopic
ratio in PNe {\it without} $^3$He measurements.  Galli et al. (1997,
hereafter GSTP) have argued that extra-mixing processes must be at work
in more than 70\% of low-mass stars ($M\simless 2$~\msun) in order to
reconcile the predictions of the Galactic evolution of $^3$He with the
observational constraints.  We set out this experiment to determine the
isotopic ratio in a relatively large sample of PNe.

\section{Observations}

The observations were carried out with the IRAM 30-m telescope at Pico
Veleta (near Granada, Spain) during two observing runs in November 1996
and May 1997.  In the first one we observed the six PNe studied by
Balser et al.~(1997). The observations were made simultaneously in the
$J = 2$--1 and 1--0 lines of $^{12}$CO. The strongest emitters were
then observed in the $^{13}$CO lines.  In the second run we searched
for $^{13}$CO $J = 2$--1 and 1--0 emission in 22 objects previously
detected in $^{12}$CO by Huggins et al.~(1996).  At the rest frequency
of the $J = 2$--1 and  1--0 $^{12}$CO lines (near 115 and 230 GHz,
respectively) the telescope beamsize and efficiency are 24$\arcsec$ and
0.6 (115 GHz) and 12$\arcsec$ and 0.45 (230 GHz). The smaller beam size
at 230 GHz, together with the small scale structure usually found in
PNe and with the high intrinsic CO (2--1)/(1--0) line ratio, lead to
observed 2--1/1--0 line ratios in the range 2--5 (Bachiller et al.~1993), 
meaning that the $J = 2$--1 line is much more effective for CO
searches.  The spectrometers used were two filter banks of 512 $\times$
1 MHz channels, providing a spectral resolution of 1.3 km s$^{-1}$ in
the $^{12}$CO 2--1 line, and 2.6 km s$^{-1}$ in the $^{12}$CO 1--0
line. Typical system temperatures were around 600 K at 115 GHz and
800--1000 K at 230 GHz.  The spectra were calibrated with the standard
chopper wheel method and are reported here in units of main-beam
brightness temperature ($T_{\rm MB}$).

\section{Measuring the carbon isotopic ratio from millimeter wave 
observations}

Molecular line observations at mm-wavelengths provide a powerful method
to estimate the \crat ratio in PNe. In particular, the
$^{12}$CO/$^{13}$CO ratio should faithfully reflect the atomic \crat
ratio, since the mechanisms which could alter the $^{12}$CO/$^{13}$CO
ratio are not expected to be at work in PNe.  In fact, the kinetic
temperature in PN envelopes (25--50 K) is high enough that 
isotopic fractionation should not operate. Also, selective
photodissociation is expected to be compensated by the isotope exchange
reaction $^{12}$CO+$^{13}$C$^+ \rightarrow ^{13}$CO+C$^+$ which is
faster than the $^{13}$CO photodestruction in PN envelopes (e.g. Likkel
et al.~1988).  However, although the $J = 1$--0 and $J = 2$--1 lines of
$^{12}$CO have been extensively observed in PNe (e.g. Huggins et al.~1996), 
very few observations of the $^{13}$CO lines are available and
the value of the isotopic ratio is presently poorly known.

In order to estimate the $^{12}$CO/$^{13}$CO isotopic ratio, one needs
to make a number of approximations. First, we assume that the emitting
regions fill the antenna beams in the lines of both molecules, or that
the filling factor is the same (in the case of an extended clumpy
medium). Second, we assume that the rotational levels are thermalized
at a representative uniform temperature of 25 K (see e.g. Bachiller et
al.~1997). Thermalization is indeed a reasonable assumption for
$^{12}$CO and $^{13}$CO, since the dipole moment is quite small (about
0.1 Debye).  Third, if we assume that the emission is optically thin
for both the $^{12}$CO and $^{13}$CO lines, then the
$^{12}$CO/$^{13}$CO column density ratio is given by the ratio of the
integrated intensities. We will discuss below the uncertainties
introduced by this approach in the derived isotopic ratios.

\section{Results}

\subsection{PNe with $^3$He measurements}

\begin{figure*}
\resizebox{\hsize}{!}{\includegraphics[angle=-90,width=8cm]{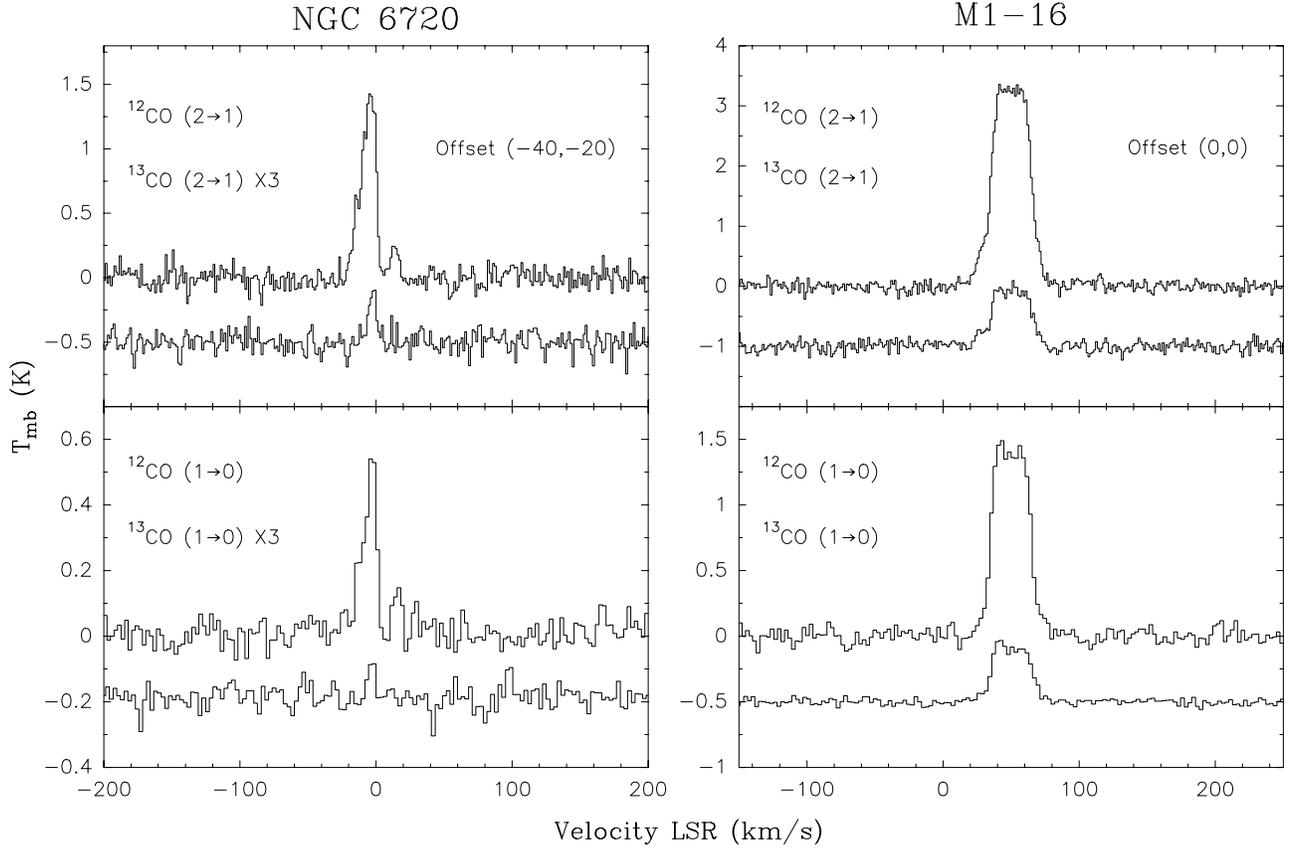}}
\caption[]{CO emission in PNe. Two examples are shown from the sample of PNe
with ({\it left}) and without ({\it right}) $^3$He measurements.
The offset in arcsec from the central PN is also indicated.
The stronger $J=2$--1 transitions are shown in the upper panels, while the
$J=1$--0 transitions in the bottom panels. The isotopic lines are detected
in both cases.}
\end{figure*}

The results of the observations are given in Table~1 where we list the
PN name, the intensities $I_{21}$ of the $^{12}$CO and $^{13}$CO 
$J=2$--1 lines and the derived isotopic ratio.  The sample of the six PNe
studied in $^3$He shows little emission in CO: with the exception of
NGC~6720, the values of I$_{21}$ represent upper limits to the
intensity and have been estimated from the line widths deduced from the
expansion velocities listed in the catalog of Acker et al.~(1992).

Bachiller et al.~(1989) detected an extended molecular envelope in
NGC~6720, the Ring nebula. The CO emission reveals a clumpy ring, resembling 
to that of the ionized nebula.  We observed the $J=2$--1 and $1$--0 lines
of $^{13}$CO in three positions and detected emission in all
cases.  Fig.~1 shows the spectra of the four CO and $^{13}$CO
lines observed toward the strongest peak in the molecular envelope
(offset [--40{\arcsec}, --20{\arcsec}] from the central position).  
The measured isotopic ratio toward this position, where the S/N is the
highest, is \crat $=22$. This value is in
agreement with a previous estimate of Bachiller et al.~(1997) and with the
values obtained toward the two positions with lower S/N spectra. Thus,
our data provide no evidence for variations of the isotopic ratio across the
nebula.

We have also detected a line around the $^{12}$CO $J = 1$--0 frequency
in the central position of NGC~6543, but we failed to detect the $J =
2$--1 line at relatively low levels. This indicates that the line near
the $J=1$--0 frequency is probably not due to CO. It is interesting to
recall that the H38$\alpha$ recombination line is only separated by 3
MHz (7.8 km~s$^{-1}$) from the $^{12}$CO $J = 1$--0 line. As discussed in
Bachiller et al.~(1992), the H38$\alpha$ line can dominate the
emission around the $^{12}$CO $J = 1$--0 frequency in some nebulae with
little or no molecular gas. We believe that this is the case in the
central position of NGC~6543.

\begin{figure}
\resizebox{\hsize}{!}{\includegraphics[width=8cm]{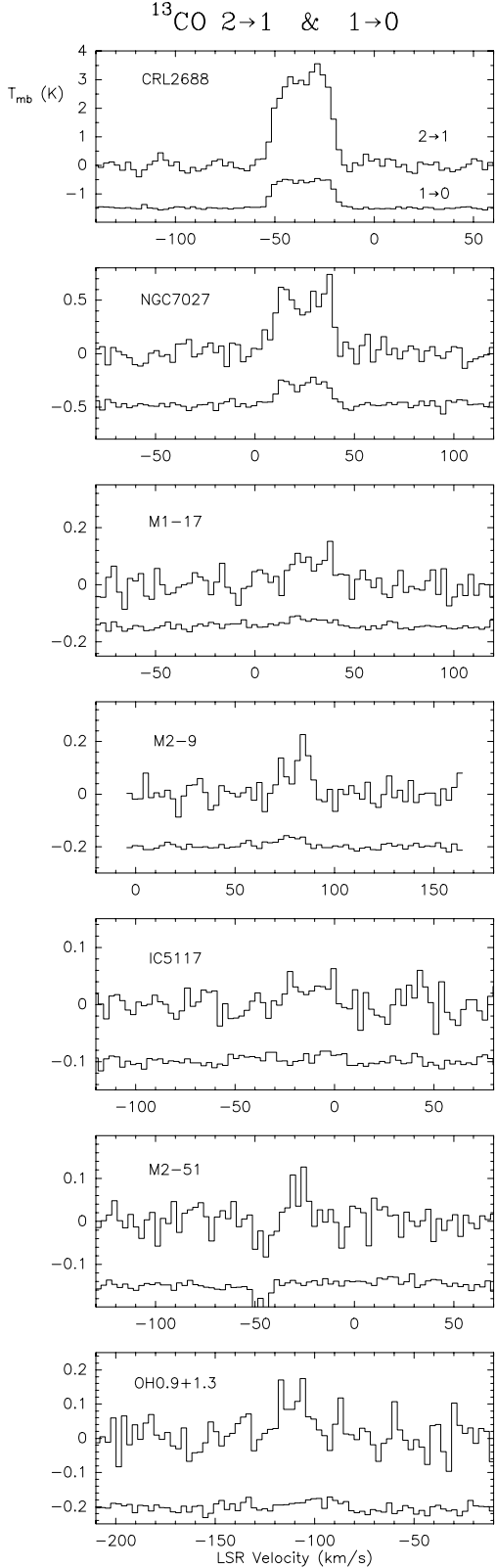}}
\caption[]{$^{13}$CO $J=2$--1 and 1--0 spectra observed toward the central 
position of the PN identified in the upper left corner of each panel.} 
\end{figure}

\begin{table}
\caption[]{CO results for PNe observed in $^3$He}
\begin{tabular}{lrrr}\hline
 PN Name  &   $I_{21}$($^{12}$CO) &  
	$I_{21}$($^{13}$CO) & $^{12}$C/$^{13}$C \\
 &  (K km s$^{-1}$)& (K km s$^{-1}$) & \\
\hline
IC~289    & $<$0.90 &     &     \\
NGC~3242  & $<$0.16 &     &     \\
NGC~6543  & $<$0.19 &     &     \\
NGC~6720  & 20.0 & 0.9 & 22  \\
NGC~7009  & $<$0.12 &     &     \\
NGC~7662  & $<$0.12 &     &     \\
\hline
\end{tabular}
\end{table}

\begin{table}
\caption[]{CO results for PNe not observed in $^3$He}
\begin{tabular}{lrrrr}\hline
 PN Name  & $I_{21}$($^{12}$CO) &  
	$I_{21}$($^{13}$CO) & $^{12}$C/$^{13}$C \\
 & (K km s$^{-1}$) & (K km s$^{-1}$) & \\
\hline
& & & & \\
NGC~7027   &  448.7  & 17.3      & 25$^\dagger$ \\
NGC~2346   &  14.2   & 0.58      & 23~   \\
NGC~7293   &  17.5   & 1.8       & 9~  \\
NGC~6781   &  28.4   & 1.72      & 20~   \\
M~1-16     &  109.0  & 32.3      & 3$^\dagger$  \\
M~1-17     &  66.2   & 3.2       & 22~   \\
M~2-9      &   3.3   & 2.0       & 2$^\dagger$  \\
M~2-51     &  33.1   & 2.3       & 15~   \\
M~4-9      &  32.2   & 1.8       & 18~   \\
CRL~2688   &  312.5  & 112       & 3$^\dagger$ \\
CRL~618    &  182.2  & 41.2      & 4$^\dagger$  \\
OH~09$+$1.3&  5.0    & 2.5       & 2$^\dagger$  \\
IC~5117    &  19.6   & 1.4       & 14~   \\
NGC~7008   &  1.8    & $\le$0.15 & $\ge$12 \\
NGC~6853   &  6.9    & $\le$0.15 & $\ge$46 \\
M~1-13     &  22.2   & $\le$1.5  & $\ge$15 \\
BD$+30^\circ 3639$ & 4.7 & $\le$1.1 & $\ge$4 \\
\hline
\end{tabular}

\noindent
$^\dagger$~lower limit 

\end{table}

\subsection{PNe without $^3$He measurements}

The parameters of the detected sources and the derived isotopic ratios
are listed in Table 2. We have also included in this table data for
NGC~2346, NGC~7293, NGC~6781, M~4-9, and CRL~618 from previous
observations by Bachiller et al.~(1989, 1997).  The spectra of the
remaining PNe are shown in Fig.~1 (for M~1-16) and in Fig.~2. In
four cases (NGC~7008, NGC~6853, M~1-13, BD$+30^\circ 3639$) 
we obtained tentative detections in the $^{13}$CO lines, 
or only crude upper limits could be derived.

The assumption of optically thin emission used to determine the
isotopic ratio is likely to be accurate in the case of evolved nebulae
like the Ring, the Helix, NGC~2346, etc. In fact,
Large-Velocity-Gradient (LVG) models confirm that the $^{12}$CO and
$^{13}$CO line emission is optically thin in these cases. On the other
hand, the same assumption is not appropriate for the CO lines in young
objects such as CRL~2688, CRL~618, NGC~7027, and M~1-16. In such cases,
the derived CO column densities and $^{12}$CO/$^{13}$CO column density
ratios represent only approximate lower limits.

One could also have weak emission arising from small optically thick
clumps very diluted within the $^{12}$CO and $^{13}$CO beams.  This
could be the case in some compact CO envelopes such as the Butterfly
nebula, M~2-9. The CO in M~2-9 is concentrated in an expanding clumpy
ring which has a mean diameter of 6 arcsec (Zweigle et al.~1997).
Individual clumps in this ring have sizes $<4$ arcsec. The weakness of
the $^{12}$CO and $^{13}$CO lines we observe could be due to the
important dilution of such small clumps within the 30-m beam.  The
clumps could be optically thick in CO, and the reported
$^{12}$CO/$^{13}$CO intensity ratio would just represent a lower limit
to the abundance ratio in these compact PNe.

In the case of extended PNe (e.g. NGC~6720, NGC~2346, NGC~7293), 
the assumption of optically thin CO 2--1
emission is based on detailed studies of individual nebulae which 
show that the intrinsic 2--1/1--0 line intensity ratio is typically $>$2
(Bachiller et al. 1989, 1993; Huggins et al. 1996). Such high ratios
imply that the emission is optically thin, and that the excitation
temperature is T$_{\rm ex} \simgreat$10 K. 
In this approximation, and assuming
homogeneous excitation conditions along the line of sight, the column
density in the $J$=2 level is proportional to the observed 2--1 line intensity.
Moreover, the relative population of the $J$=2 level is quite
insensitive to the value of the excitation temperature for the typical
conditions of PNe. 
The total CO column densities determined in
this way are correct for T$_{ex}$ in the range from 7 to 77 K, and
are within a factor of 2 over the range from 5 to 150 K, which should cover
most PNe. In any event, because of the similar dipole moments, the excitation 
temperature should be the same for both CO and $^{13}$CO. 
Then, if the excitation conditions do not vary along the line of sight, the 
isotopic ratios are independent of T$_{ex}$. 

The major source of uncertainty for these extended nebulae is thus related
to the observational procedures. The calibration of our observations is
accurate within 20{\%}, but small differences in the filling factor 
for the different spectral lines 
could increase the uncertainty of the measured isotopic ratios up to
a factor $\sim$1.5. As Table 2 shows, 
in the extended PNe where $^{13}$CO is well detected (NGC~6720, NGC~2346, 
NGC~7293, NGC~6781, M~4-9, and M~2-51) the values of the isotopic ratios 
are in the range 9 to 23, to within a factor of $\sim$ 1.5. 
Such values
are thus significantly lower than the Solar System value of 89, and
appear to be consistent with those derived in AGB stars, as we shall discuss
in Sect.~6.1.

\section{Mass estimate of the progenitors of the PNe}

In order to compare the observed isotopic ratios with the predictions
of stellar evolutionary models, we must estimate the mass of the
progenitors stars of the PNe.  For the present samples, we have
followed the procedure adopted in GSTP, consisting of the following
steps.

(1) We assign to each nebula the best measured distance (or an average
of available distances), the H$\beta$ flux, the He{\sc ii} flux at
4686~\AA, the angular size, and the $B$ and $V$ stellar magnitudes.
Using these parameters, we calculate the effective temperature and
luminosity via the Zanstra method (Kaler~1983).

(2) By locating the central star in the $\log T_{\rm eff}$--$\log L$ plane, 
we derive its mass ($M_{\rm CS}$) from comparison with a set of
evolutionary tracks (Stanghellini \& Renzini 1993).

(3) Using the initial mass--final mass relation, we compute the progenitor
mass, that is the stellar mass on the main sequence ($M_{\rm MS}$).

The stellar properties adopted for the PNe detected in $^{13}$CO and
the derived values of the progenitor mass are given in Table~3.  
Details on the individual objects are given in the Appendix.

\begin{table*}
\caption[]{Properties of individual PNe detected in $^{13}$CO} 
\begin{tabular}{rlllccr}\hline
\# & PN Name & Log $T_{\rm eff}$ & Log $L$ & $D$ & $M_{\rm CS}$ & $M_{\rm MS}$ \\
   &  & (K) & ($L_\odot$) & (kpc)  & ($M_\odot$) & ($M_\odot$) \\
\hline
& & & & & & \\
1 & NGC~7027  & $5.28\pm 0.02$ & $3.66\pm 0.03$ & 0.88 & $0.67\pm 0.03$ & $2.6$ \\
2 & NGC~2346  & $4.76\pm 0.01$ & $3.80\pm 0.04$ & 0.90 & $0.63\pm 0.02$ & $2.3$ \\ 
3 & NGC~7293  & $5.03\pm 0.01$ & $2.03\pm 0.03$ & 0.28 & $0.60\pm 0.03$ & $2.6$ \\
4 & NGC~6781  & $4.99\pm 0.03$ & $2.94\pm 0.09$ & 1.57 & $0.55\pm 0.02$ & $1.0$ \\
5 & M~1-16    & $4.91\pm 0.03$ & $3.36\pm 0.14$ & 5.43 & $0.56\pm 0.02$ & $1.0$ \\
6 & M~1-17    & $4.97\pm 0.03$ & $2.89\pm 0.11$ & 7.36 & $0.55\pm 0.05$ & $1.0$ \\ 
7 & M~2-9     & $>4.65$         & $>2.54$         & 1.32 & $(^\ast)$      & $>2$  \\
8 & M~2-51    & $5.06\pm 0.06$ & $2.20\pm 0.15$ & 1.92 & $0.63\pm 0.09$ & $2.3$ \\
9 & M~4-9     & $>4.85$         & $>0.18$         & 1.71 & $(^\ast)$      &       \\ 
10 & CRL~2688  &                  &                  &      &                & $>1.3$\\ 
11 & CRL~618  & $>4.48$   & $>1.11$   & 2.81 & $(^\ast)$      &       \\
12 & OH~09$+$1.3 &                &                  & 8.0  &                &       \\
13 & IC~5117   & $4.95\pm 0.03$ & $3.31\pm 0.12$ & 2.10 & $0.56\pm 0.02$ & $1.0$ \\
\hline
\end{tabular}

\noindent 
$(^\ast)$~no He{\sc ii}, no Zanstra analysis possible
\normalsize
\end{table*}

Let us examine the uncertainty involved in the final mass
calculations.  Estimates of the stellar temperature and luminosity
given in Table~3 are affected by errors in magnitudes, fluxes,
diameters and extinctions. However, these quantities are usually
determined with good accuracy ($\sim 5-20\%$), so that the uncertainty
in the derived mass of the central stars does not exceed $\sim 15$\%,
or $\sim 0.02$~\msun. The values given in the table do not include
errors on the distances to the PN, which can be intrinsically high (up
to 50\%) but are difficult to estimate on an individual basis.

To infer the main sequence masses, we have used the initial mass--final
mass relation given by Hervig~(1996). This relation differs from that
of Weidemann~(1987) adopted in GSTP. We preferred Hervig's prescription
since it is based on reliable observations of cluster white dwarfs,
although the formal errors on the final mass are still substantial, and
can amount to about 0.1 $M_\odot$. Since we derive initial masses from
final masses, the errors on the former quantity can be even larger.
Quantitatively, we assign a formal error to the main sequence mass of
$\Delta M_{\rm MS}\simeq 1.5$~\msun for low values of the initial mass
($M_{\rm MS}<2$~\msun), and a smaller error ($\Delta M_{\rm MS}\simeq
0.75$~\msun) for higher masses. This difference is due to
the change of the slope of the
initial mass - final mass relation at about 2 \msun: smaller masses
are more sensitive to the adopted relation, and the uncertainty
is correspondingly larger.

\section{Implications}

\subsection{Implications on stellar nucleosynthesis}

\begin{figure}
\resizebox{\hsize}{!}{\includegraphics{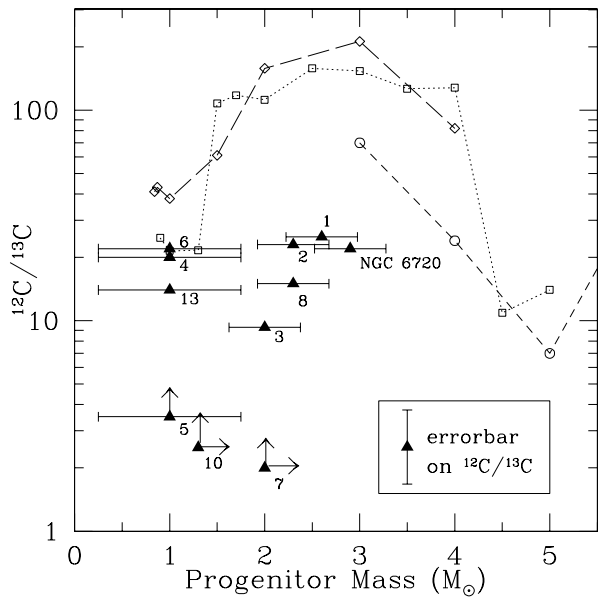}}
\caption[]{\crat versus progenitor mass of the PNe detected in CO.
The filled triangles are the measured PNe, labelled as in Table 3.
The uncertainty on the mass
is described in the text. The error bar on the derived \crat ratio is
shown in the lower right corner. The three curves are the
predicted \crat ratios in the {\it ejecta} of stars at the tip of the
asymptotic giant branch (AGB). The {\it dotted} curve is from van den
Hoek \& Groenewegen (1997), the {\it dashed} curve is from Forestini \&
Charbonnel (1997), and the {\it long-dashed} curve is from Marigo
(1998). All curves are for solar metallicity.} 
\end{figure}

How do we interpret our results on the \crat\ ratios in the framework of
stellar nucleosynthesis?  To help answering this question, we combine 
the information provided by the observed \crat\ isotopic ratios
with the mass estimates of the progenitors of the PNe, and with the
predictions of some representative stellar nucleosynthesis models.

Since the formation of a PN takes place at the end of the AGB phase, the
significant comparison is between the observed abundances and
those predicted for the stellar ejecta at the AGB tip. 
Unfortunately, no stellar nucleosynthesis models
up to these late phases are available in the literature for stars
experiencing deep-mixing. For example, the calculations of Boothroyd
\& Sackmann (1999) allow to derive the mass dependence of the \crat\
ratio on the stellar surface at the tip of the red giant branch both
for the standard case and in the presence of extra-mixing.
The standard models indicate that the isotopic ratio has
approximately a constant value of 20--23 between
$M=2$ and 4~\msun, and then increases steadily at lower masses up to about 
28--30 with a small dependence on the stellar metallicity. 
In the case of extra
mixing, the \crat ratio displays a sharp drop below $\sim 2$~\msun,
reaching values of 5--10 at $\sim$1 M$_\odot$.
Similar results have also been obtained by Charbonnel (1994) and
Denissenkov \& Weiss (1996).

In Fig.~3 we show the distribution of the measured \crat ratios
in 11 PNe, together with the theoretical values in the ejecta of stars
at the tip of the AGB phase. The different curves refer to the models
computed in the standard case with no mixing and solar metallicity 
by FC, HG, and Marigo (1998). 
Fig. 3 indicates that over the mass interval between 1.5 and 4 M$_\odot$
the predictions of the HG and Marigo models are in good agreement with
each other and indicate
a roughly constant value of \crat$\sim$100. The models of FC show 
significantly lower ratios at 3 and 4 $M_\odot$, but have the same 
qualitative behavior. These results are well understood in terms
of the nucleosynthesis occurring during the thermally pulsating AGB phase.
In fact, a major phase of $^{12}$C enrichment of the convective envelope
results from the penetration of the convective tongue during thermal pulses
(see the discussion in, e.g., FC and HG). At the same time, 
$^{13}$C is partially 
burnt through the $^{13}$C($\alpha$,n) reaction at the bottom of the
inter-shell region during the inter-pulse phase, as first suggested by 
Straniero et al. (1995). The combination of these two effects accounts for
the high \crat ratio in this mass range. 
On the other hand, more massive AGB stars experience hot bottom burning
that progressively leads the \crat\ ratio close to its equilibrium value
($\sim$4-5).

Despite the uncertainties in our measurements, 
it appears that the standard models
produce \crat in excess of the values observed in our objects.  The
only exception is for the two lowest mass stars (0.9 and 1.0 $M_\odot$)
computed by HG which agree with the ratios measured in M1-17 and NGC~6781.  
Although Fig. 3 indicates a marked discrepancy between observed and 
theoretical values, one should be aware of the 
sensitivity of the predicted yields on the model assumptions.
In particular, the isotopic ratios (not only \crat) 
depend rather strongly on the adopted mass loss rate during the AGB phase, 
on the third dredge-up efficiency, and on stellar metallicity. 
Their combined effects 
have been thoroughly discussed by FC and HG and need not to be repeated here.
In general, a shorter phase of mass loss and/or lower mass loss
rates tend to decrease the surface \crat\ ratio. However, a more efficient
convective penetration results in a higher pollution of the 
envelope from pulse to pulse and a higher \crat\ ratio. Finally,
stars with initial metallicities lower than solar yield higher \crat ratios.

Returning to Fig. 3, we see that
several PNe lie in the region of the diagram with low \crat and
low mass. For these objects we should expect the standard models to
become inaccurate insofar as they neglect the possible occurrence
of mixing mechanisms of non-convective origin which can alter 
the composition of the stellar ejecta.  
Low values of the carbon isotopic ratio have also been measured in
field population II stars and in globular cluster giant (e.g. Sneden et al.
1986; Gilroy \& Brown 1991; Pilachowski et al. 1997) and have
provided the motivation to introduce extra-mixing processes in the
standard evolution (e.g. Charbonnel~1995).  
Recently, Charbonnel \& do
Nascimento~(1998) find that more than 90\% of a sample of 191 field and
cluster red giants presents carbon isotopic ratios inconsistent with
those predicted by standard nucleosynthesis.

The observational situation in AGB stars is less clear, because
of the extreme sensitivity of the determination of isotopic ratios on
excitation temperatures and model atmospheres.  For example, Greaves \&
Holland (1997) have measured a \crat ratio varying between 12 and 36 in a
sample of 9 carbon stars with high mass-loss rates ($\sim 10^{-5}$
M$_{\sun}$~yr$^{-1}$), in accordance with similar results previously obtained
by Wannier \& Sahai (1987) on seven other carbon stars.  On the other hand,
Lambert et al. (1986) measured the \crat\ ratio in 30 cool carbon stars and
found values between 30 and 70.  More recently, Ohnaka \& Tsuji (1996) report
ratios between 20 and 30 for 24 N-type carbon stars in common with the sample
of Lambert et al.  (1986). These
values are about a factor of 2 smaller than those derived by Lambert et al.
(see however the discussion in de Laverny \& Gustafsson 1998 and 
Ohnaka \& Tsuji 1998), but are consistent with those found for our PN sample.  If these
samples are representative of the population of both carbon stars and PNe,
the low \crat ratios provide strong indication that most stars should
experience some extra-mixing and deplete $^{12}$C with respect to $^{13}$C.
According to this conjecture, it appears that in order to match the
observations, {\it more realistic stellar models should include additional
physical processes to avoid too large} \crat\ {\it ratios throughout the
latest stages of stellar evolution}.

\subsection{Implications on Galactic chemical evolution}

Theoretical models predict that stars with low values of \crat should
also have low $^3$He abundances.  In GSTP, we estimated that to
obtain consistency between the observed $^3$He abundances and chemical
evolution models, the majority of the Galactic PNe ($>$70\%) should
have a \crat ratio well below the standard value.  In this section we
test this suggestion using the constraint provided by the \crat
observations in PNe.  Chemical evolution models offer the adequate tool
to follow simultaneously the evolution of \crat and $^3$He over the
Galactic lifetime.

\begin{figure}
\resizebox{\hsize}{!}{\includegraphics{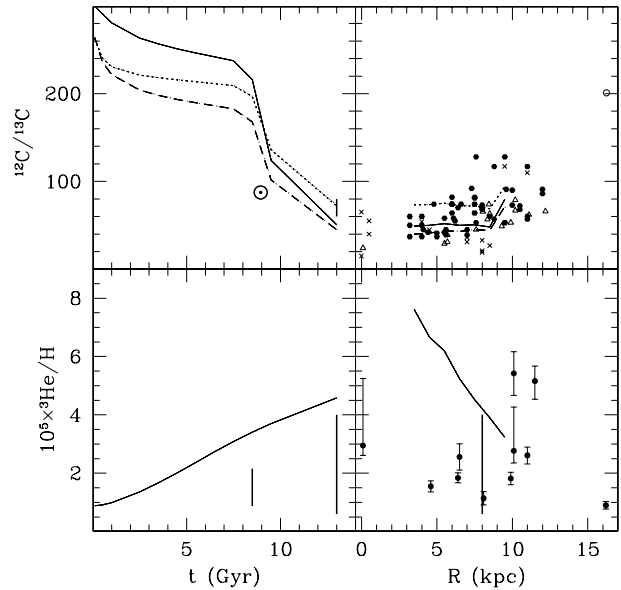}}
\caption[]{Evolution of \crat and $^3$He as a function of time for the
solar neighborhood and as a function of Galactocentric radius at the
present time. Standard case with no extra mixing. The three curves show
the results obtained using the yields for intermediate-mass stars by FC
({\it solid}), by HG ({\it dotted}), and by BS ({\it dashed}).  The
observational data for \crat are from: Anders \& Grevesse (1989);
Henkel et al.~(1980); Gardner \& Whiteoak~(1981); Henkel
et al.~(1982); Henkel et al.~(1985), Langer \&
Penzias~(1993); Wouterloot \& Brand~(1996). The data on $^3$He are from
Geiss~(1993), Gloeckler \& Geiss~(1996) and Rood et al.~(1995).  }
\end{figure}

\begin{figure}
\resizebox{\hsize}{!}{\includegraphics{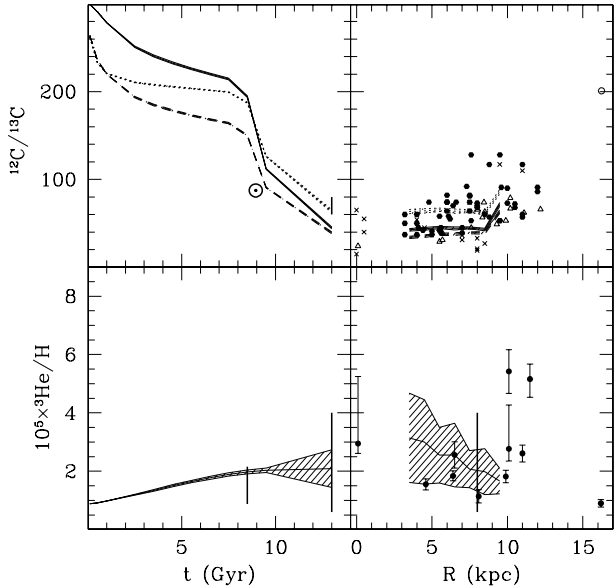}}
\caption[]{Evolution of \crat and $^3$He as a function of time for 
the solar neighborhood and as function of Galactocentric radius
at the present time. Case with mixing in 90\% of low mass stars. 
Symbols and observational data are as in Fig.~4.}
\end{figure}

In Figs. 4 and 5 we show the evolution of \crat and $^3$He as a
function of time in the solar neighborhood (Galactocentric radius
$R=8$~kpc), and as a function of $R$ at the present time ($t=13$~Gyr),
as predicted by models described by Dearborn et al.~(1996)
and Sandrelli et al.~(1998).  The two figures correspond to
different assumptions on the fraction of low-mass stars experiencing
deep mixing in the RGB phase. In both cases, the Galactic evolution of
the various isotopes has been computed assuming metallicity-dependent
yields. In the upper panels, the solid line corresponds to models
adopting the $^{12}$C and $^{13}$C yields from BS for stars in the mass
range 0.8 -- 2.5 \msun, from FC in the range 3 -- 6 \msun, and from
Woosley \& Weaver~(1995) for massive stars. The dotted and dashed lines
display the results obtained using the same yields as before, except
for stars in the mass range 3--6 \msun, where the HG and BS yields,
respectively, are adopted.  The calculations by BS are the only ones
that provide the $^{12}$C and $^{13}$C yields for both the standard and
the CBP cases. For $^3$He, the standard yields are taken from Dearborn
et al.~(1996) and the CBP yields from BS.

The behaviour of $^3$He is typical of the elements produced mainly by
low-mass, long-lived stars. $^{12}$C is a primary element produced by
stars of any mass and available for Galactic enrichment already after
the explosions of the first massive objects. $^{13}$C is mostly
secondary and its bulk abundance is due to intermediate-mass stars.
Therefore, the enrichment of this element in the ISM occurs later than
that of $^{12}$C. This delay causes the decrease of \crat visible
in Figs.~4 and 5. The positive radial gradient of \crat is also due to
the different mass intervals of stars that are producers of the carbon
isotopes, and to the higher star formation activity in the inner
Galactic regions, independent of the stellar initial mass function. All
these effects combined make the relative proportion of $^{13}$C over
$^{12}$C producers increasingly higher for decreasing Galactocentric
distances.

Fig.~4 shows the standard case with no deep mixing, whereas the
results with CBP in 90\% of stars randomly chosen in the mass range
$M\simless 2.5$~\msun\ are displayed in Fig.~5.  The hatched regions
represent the range of abundances resulting from 50 different cases for
the random sorting.  As already discussed by GSTP, the overall
enrichment of $^3$He is very sensitive to the assumptions on deep
mixing. On the contrary, the variation of \crat is marginal, since most
of $^{12}$C and $^{13}$C is produced by stars more massive than
2.5~\msun\ which are not affected by deep mixing processes. Thus,
altering the amount of $^{12}$C and $^{13}$C released by low-mass stars
leads to smaller variations in the resulting abundances.  For these
reasons, the Galactic evolution of $^3$He is an excellent indicator of
the fraction of low-mass stars which should undergo CBP, but that of
\crat is not.  From Figs.~4 and 5, one can see that the fraction
required to reproduce the observed pattern of $^3$He abundances is up
to 90\%, in agreement with the finding of Charbonnel \& do Nascimento~(1998).

As for $^{12}$C/$^{13}$C, the predicted abundance distributions with
time and with Galactocentric distance are roughly consistent with the
data regardless of the occurence of extra-mixing.  The \crat ratio
in the Sun cannot be accurately reproduced, and the radial gradient is
flatter than the observed one (as deduced from measurements in
molecular clouds), independently of the adopted percentage of stars
undergoing CBP.  Using the yields of FC and BS leads to a better fit of
the solar value and to the present distribution of \crat in the inner
Galactic regions. On the other hand, the models of HG reproduce the
local ISM ratio.  In all cases, the fraction of low-mass stars
experiencing deep mixing does not affect significantly the overall
results of the chemical evolution models. Of course, deep mixing alters
the values \crat in low-mass stars, but the Galactic evolution of
$^{12}$C and $^{13}$C is mainly governed by stars in which this process
is not expected to take place.

\section{Prospects for future observations}

One of the main goals of our study was the determination of the
isotopic ratio in NGC~3242, the best object for testing the mixing
hypothesis since the $^3$He measurements are very reliable {\it and}
the progenitor mass is in the range where extra mixing should occur
(see also the discussion in Rood et al.~1998).  However, the high
sensitivity of our observations and the fact that we searched for CO
emission across the whole circumstellar shell imply that CO is absent
in the nebula.  We therefore conclude that the isotopic ratio cannot be
determined in this important object by means of millimeter line
observations.  The lack of a molecular envelope around NGC~3242 can be
a consequence of the low-mass of the progenitor. It is well known that
the transition from the AGB phase to the PN stage is slow for low-mass
stars and therefore the gas remains exposed to ionizing and
dissociating radiation for a longer time than in the case of more
massive stars.

This behavior is consistent with the fact that the most massive object of 
our sample, NGC~6720, indeed has a rich molecular envelope.
Unfortunately, NGC~6720 has an estimated progenitor mass of $\sim 2
$~\msun, quite close to the limit where the nonstandard mixing
mechanism is expected to significantly affect the isotopic ratio.
Thus, the derived ratio is consistent with both standard and
nonstandard evolutionary models. The lesson is clear:  there is a
trade-off between the need to select low-mass PNe in order to
discriminate between theoretical models and the higher detection rate
of molecular envelopes around more massive objects. The
determination of isotopic ratios may be a tricky business, but future
observations should take up such a challenge.

The method of using mm-wave transitions is not the only one adequate
for measuring the isotopic abundance in PNe. Clegg~(1985) first pointed
out the possibility of using the CIII] multiplet near 1908~\AA~ to
obtain a direct estimate of the isotopic ratio in the {\it ionized} gas
of PNe. Recently, Clegg et al.~(1997) have successfully detected
the extremely weak isotopic line $^{13}$C 
$^3_{1/2}$P$^\circ_0$--$^1_{1/2}$S$_0$ in two PNe (plus a tentative
detection in a third one), using the HST Goddard High Resolution
Spectrograph.  The derived \crat ratios are 15$\pm$3 in NGC~3918 and
21$\pm$11 in SMC~N2.  In either case, no measurement at mm-wavelengths
has been made to independently check the derived values.  However, as
Clegg et al. point out, the two types of measurements are complementary
since they cannot be performed on the same objects: the UV transitions
require high excitation conditions, while the opposite is true for the
mm-wave lines studied by us.  Considering the negative results of the
CO search, it would be extremely important to perform the HST
observations towards NGC~3242.

\section{Conclusions}

We have observed a sample of 28 PNe in $^{12}$CO and $^{13}$CO.
We have determined the isotopic ratio in 14 objects, 9 of which are new
detections, and obtained
robust upper limits in 4 other PNe.  The observations reported in this
paper improve the available data on the \crat ratio in PNe. The results
confirm and extend earlier measurements which indicate values between 5
and 30, below those predicted by current AGB models.  According
to these calculations, the value of \crat in the ejecta of stars in the
mass range between $\sim 1.5$ and $\sim 4$~\msun\ is of the order of
100. In order to agree with the observations, we suggest that
stellar nucleosynthesis models should include the effects of
extra-mixing to avoid the overproduction of $^{12}$C with respect to
$^{13}$C not only during the RGB phase, but also in the latest phases of
evolution.  Our estimates of the mass of the progenitor stars support
this conclusion, since all PNe considered here originated from
stars with mass less than $\sim 3$~\msun. The proposed extra-mixing
mechanism is expected to be at work in this mass range (Hogan~1995,
Charbonnel~1995) and represents at present the most likely solution
to the so-called ``$^3$He problem''.

Out of the six PNe with measured $^3$He abundances, only for one
(NGC~6720) it was possible to estimate the \crat ratio. The remaining
five objects show weak or no $^{12}$CO emission. In the case of
NGC~3242 where the $^3$He measurements are the most reliable, the
absence of CO emission in the envelope precludes a direct test of the
predictions of stellar nucleosynthesis using millimeter line observations.
We have suggested an alternative method to determine the isotopic 
abundance in this critical object.

We have used chemical evolution models to follow the variation with
time and Galactocentric distance of the $^3$He abundance and \crat
ratio.  Whereas the model predictions for $^3$He are inconsistent with
the observed abundance in the Solar System and in the interstellar
medium unless extra-mixing is operative in at least $90\%$ of low-mass
stars, the evolution of \crat is remarkably insensitive to the effects
of extra-mixing.  The Galactic evolution of $^{12}$C and $^{13}$C is
mainly governed by intermediate and massive stars, in which
extra-mixing is not expected to occur.

\begin{acknowledgements}
We would like to thank M.~P\'erez
Guti\'errez for his help with part of the observations. We acknowledge
R. Rood, T. Bania and D. Balser for numerous and fruitful discussions
on the $^3$He measurements. We thank the referee for useful comments which
improved the content of the paper.  The research of F.P., M.T., and D.G
has been supported by grant COFIN98-MURST at the Osservatorio di Arcetri.
R.B. acknowledges partial funding from Spanish DGES grant PB96-104.
\end{acknowledgements}

\appendix
\section{Notes on individual PNe}

Here, we provide detailed information on the physical properties 
of the PNe of our sample.  Unless otherwise
noted, $B$ and $V$ magnitudes are from the Acker et al.~(1992)
catalogue. Fluxes, extinction constants, and diameters are from Cahn
et al. (1992, hereafter CKS).

\noindent
{\bf NGC 7027}: The Zanstra analysis yields $\log T_{\rm eff}=5.28\pm
0.02$ and $\log L/L_{\odot}=3.66\pm 0.03$.  We have used the
expansion distance from the radio image of Masson~(1989). The derived
mass of the central star is $M_{\rm CS}=0.67$~\msun.

\noindent
{\bf NGC~2346}: The distance, $D=0.9$~kpc, is derived from the average
extinction distance (Pottasch~1996).  We obtain $\log T_{\rm
eff}=4.76\pm 0.01$ and $\log L/L_{\odot}=3.80\pm 0.04$. The mass of
the central star results $M_{\rm CS}=0.63$~\msun.

\noindent
{\bf NGC~7293}: Many distance measurements are available for this
nebula.  Pottasch~(1996) quotes a very uncertain (error up to 100\%)
parallax distance of $D=0.19$~kpc.  A more reliable value of
$D=0.28$~kpc comes from averaging several individual distances
(Pottasch~1996).  We obtain $\log T_{\rm eff}=5.03\pm 0.01$ and $\log
L/L_{\odot}=1.80\pm 0.03$ or $2.03\pm 0.03$, using the parallax and
average distances, respectively.  The corresponding central star masses
are $M_{\rm CS}=0.675$ and 0.60~\msun. The progenitor masses are 2.6
and 2.0~\msun. Given the large error on the parallax, we prefer the
latter value for the initial mass of NGC~7293.

\noindent
{\bf NGC~6781}: We obtain $\log T_{\rm eff}=4.99\pm 0.03$. By using
the statistical (CKS) and the average of individual (Acker et al.~1992)
distances, we obtain respectively $\log L/L_{\odot}=2.24\pm 0.09$ and
$2.94 \pm 0.09$. In both cases, the mass of the central star
coincides with the evolutionary track for $M_{\rm CS}=0.55$~\msun.

\noindent
{\bf M~1-16}: By using $D=5.45$~kpc (CKS method with a
newly measured diameter), we find $\log L/L_{\odot}=3.36\pm 0.14$ and
$\log T_{\rm eff}=4.91\pm 0.03$. The inferred mass of the central
star is $M_{\rm CS}=0.56$~\msun.

\noindent
{\bf M~1-17}: With a distance of $D=7.36$~kpc (from CKS
method), we obtain $\log T_{\rm eff}=4.97\pm 0.03$ and $\log
L/L_{\odot}=2.89\pm 0.11$. The location on the HR diagram is just
below the 0.55~\msun.

\noindent
{\bf M~2-9}: We calculate the distance with the method of CKS,
but we evaluate the effective diameter from the $H\alpha$
image ($\theta=34.75$~arcsec).  The new distance to this nebula is
$D=1.32$~kpc.  No He{\sc ii} flux has been detected. Thus, the Zanstra
analysis gives lower limits to the temperature and luminosity.  We
obtain $\log T_{\rm eff}>4.65$ and $\log L/L_{\odot}>2.54$ from the
hydrogen recombination lines. To date, nothing has been published on
this nebula to set better constraints on the mass of the central star.

\noindent
{\bf M~2-51}: The Zanstra analysis gives $\log T_{\rm eff}=5.06\pm
0.06$ and $\log L/L_{\odot}=2.20\pm 0.15$. The luminosity is
estimated using a distance $D=1.92$~kpc, derived with the CKS
method and a new estimate of the diameter. The resulting final mass is
$M_{\rm CS}=0.63$~\msun.

\noindent
{\bf M~4-9}: The distance evaluated with the newly measured angular
size based on the H$\alpha$ image by Schwarzet al.~(1992)
is $D=1.71$~kpc. For the central star, only a photographic
magnitude exists. Thus, the resulting Zanstra analysis is quite
uncertain. The $H_{\beta}$ flux is from Acker \etal. The He{\sc ii}
flux has never been measured. We obtain $\log T_{\rm eff}>4.85$ and
$\log L/L_{\odot}>0.18$. A value of the mass of the central star
cannot be derived.

\noindent
{\bf CRL~2688}: Catalogued as a ``possible planetary nebula", the
famous Egg nebula is actually a proto-PN. The standard analysis via
Zanstra method is not feasible, as the nebula is still very thick
to optical radiation. There are two mass estimates in the literature,
$M_{\rm MS}=1.3$ or 2.7~\msun respectively, which depend on the
assumed mass loss rate (Sahai et al.~1998).

\noindent
{\bf CRL~618}: The He{\sc ii} flux is not available. Therefore, we
cannot derive a Zanstra temperature. Hydrogen recombination lines give
$\log T_{\rm eff}>4.48$ and a luminosity of $\log
L/L_{\odot}>1.11$, if $D=2.81$~kpc is used.  This distance has
been derived with the CKS method and diameter measured by
Manchado et al.~(1996). No mass determination is possible.

\noindent
{\bf OH~09+1.3}: Another proto-PN, with no published mass to date.

\noindent
{\bf IC~5117}: The distance is the average of the extinction distances 
quoted in Acker et al.~(1992). We calculate $\log T_{\rm eff}=4.95\pm 0.03$
and $\log L/L_{\odot}=3.31\pm 0.12$. The derived mass of
the central star is $M_{\rm CS}=0.56$~\msun.

\noindent
{\bf M~1-59}: Since no stellar magnitudes are available for this
object, we calculate the effective temperature with the ``crossover"
method (Kaler~1983). We use the statistical distance $D=2.50$~kpc from
CKS, based on a new diameter measured by Manchado et al.~(1996) and
obtain $\log T_{\rm eff}=5.07\pm 0.01$ and $\log L/L_{\odot}=1.37\pm
0.02$. The mass of the central star mass is $M_{\rm CS}=0.85$~\msun.

\noindent
{\bf NGC~6853}: The parallax distance quoted in Pottasch~(1996) is of
the best quality, so we can confidently use this value to evaluate the
luminosity. From the Zanstra analysis we derive $\log T_{\rm
eff}=5.14\pm 0.03$ and $\log L/L_{\odot}=2.52\pm 0.07$, and a
central mass $M_{\rm CS}=0.67$~\msun

\noindent
{\bf BD+30$^\circ$3639}: The He{\sc ii} flux is not detected. Thus, we
obtain lower limits to the luminosity and temperature: $\log T_{\rm
eff}>4.68$ and $\log L/L_{\odot}>2.70$, using the hydrogen
recombination lines and the average of extinction distances from Acker et al.~(1992).
No mass determination is possible.

\noindent
{\bf NGC~7008}: We calculate $\log T_{\rm eff}=4.98\pm 0.01$ and
$\log L/L_{\odot}=3.76\pm 0.13$. The distance used is the average of
the extinction distances quoted in Acker et al.~(1992).  The derived
central mass is $M_{\rm CS}=0.63$~\msun.

\noindent
{\bf M~1-7}: The Zanstra analysis gives $\log T_{\rm eff}=5.05\pm
0.04$ and $\log L/L_{\odot}=2.43\pm 0.15$ when using $D=3.47$~kpc,
as derived from the diameter of Manchado et al.~(1996) and the method of CKS.
The mass of the central star is $M_{\rm CS}=0.56$~\msun.

\noindent
{\bf M~1-13}: Since no stellar magnitudes are available for this
object, we calculate the effective temperature with the ``crossover"
method (Kaler~1983). We use the statistical distance $D=5.32$~kpc from
CKS to obtain $\log T_{\rm eff}=5.35\pm 0.03$ and $\log
L/L_{\odot}=3.70\pm 0.26$. The central star mass is $M_{\rm CS}=0.71$~\msun.

{}


\begin{thebibliography}{}

\bibitem[1992]{A92}
Acker A., Ochsenbein F., Stenholm B., et al.
1992, Strasbourg--ESO Catalogue of Galactic planetary nebulae (Garching: ESO)

\bibitem[1989]{AG89}
Anders E., Grevesse N. 1989, Geochim. Cosmochim. Acta 53, 197

\bibitem[1989]{Bac89}
Bachiller R., Bujarrabal V., Mart\'{\i}n-Pintado J., G\'omez-Gonz\'alez 
J. 1989, A\&A 218, 252

\bibitem[1992]{Bac92}
Bachiller R., Huggins P.J., Mart\'{\i}n-Pintado J., Cox P. 1992, 
A\&A 256, 231

\bibitem[1993]{Bac93}
Bachiller R., Huggins P.J., Cox P., Forveille T. 1993, A\&A 267, 177

\bibitem[1997]{Bac97}
Bachiller R., Forveille T., Huggins P.J., Cox P. 1997, A\&A 324, 1123

\bibitem[1994]{BBBRW}
Balser D.S., Bania T.M., Brockway C.J., Rood R.T., Wilson T.L. 1994, ApJ 
430, 667

\bibitem[1997]{BBRT97}
Balser D.S., Bania T.M., Rood R.T., Wilson T.L. 1997, ApJ 483, 320

\bibitem[1998]{Ba98}
Balser D.S., Bania T.M., Rood R.T., Wilson T.L. 1998, ApJ 510, 759

\bibitem[1999]{BS99}
Boothroyd A.I., Sackmann I.-J. 1999, ApJ 510, 232 (BS)

\bibitem[1992]{cahn}
Cahn J.H., Kaler J.B., Stanghellini L. 1992, A\&AS 94, 399 (CKS)

\bibitem[1994]{Ch94}
Charbonnel C. 1994, A\&A 282, 811

\bibitem[1995]{Ch95}
Charbonnel C. 1995, ApJ 453, L41

\bibitem[1998]{CN98}
Charbonnel R., do Nascimento J.D.Jr, 1998, A\&A 336, 915 

\bibitem[1985]{C85}
Clegg R.E.S. 1985, in Production and Distribution of C, N, O Elements,
eds. J. Danziger, Matteucci, F., Kjar, K. (Munich: ESO), p.261

\bibitem[1997]{CSWN97}
Clegg R.E.S., Storey P.J., Walsh J.R., Neale L. 1997,  MNRAS 284, 348

\bibitem[1992]{co92}
Cox P., Omont A., Huggins P.J., Bachiller R., Forveille T. 1992,
A\&A 266, 420

\bibitem[1996]{DST96}
Dearborn D.S.P., Steigman G., Tosi M. 1996, ApJ 465, 887

\bibitem[1996]{DW96}
Denissenkov P.A., Weiss A. 1996, A\&A 308, 773

\bibitem[1997]{FC97}
Forestini M., Charbonnel C. 1997, A\&AS 123, 241 (FC)

\bibitem[1995]{GPFP95}
Galli D., Palla F., Ferrini F., Penco U. 1995, ApJ 443, 536

\bibitem[1997]{GSTP97}
Galli D., Stanghellini L., Tosi M., Palla F. 1997, ApJ 477, 218 (GSTP)

\bibitem[1981]{GW81}
Gardner F.F., Whiteoak J.B., 1981, MNRAS 37, 1981

\bibitem[1993]{G93}
Geiss J. 1993, in Origin and evolution of the elements, eds. N. Prantzos, 
E. Vangioni-Flam, M. Cass\'e (Cambridge: CUP), p.89

\bibitem[1991]{G91}
Gilroy K.K., Brown J.A. 1991, ApJ 371, 578

\bibitem[1993]{GG96}
Gloeckler G., Geiss J. 1996, Nat 381, 210

\bibitem[1997]{GH97}
Greaves J.S., Holland W.S. 1997, A\&A 327, 342

\bibitem[1980]{Hal80}
Henkel C., Walmsley C.M., Wilson T.L. 1980, A\&A 82, 41

\bibitem[1982]{Hal82}
Henkel C., Wilson T.L., Bieging L. 1982, A\&A 109, 344

\bibitem[1985]{Hal85}
Henkel C., Gusten R., Gardner F.F. 1985, MNRAS 143, 148

\bibitem[1995]{Her95}
Hervig G. 1996, in Stellar Evolution: what should be done, 32$^{\rm nd}$
Li\`ege Int. Astroph. Coll., eds. A. Noels, N. Grevesse (Univ. of Li\`ege), 
p.441

\bibitem[1195]{Ho95}
Hogan C.J. 1995, ApJ 441, L17

\bibitem[1996]{HBCF96}
Huggins P.J., Bachiller R., Cox P., Forveille T. 1996, A\&A 315, 284

\bibitem[1983]{K83}
Kaler J. B. 1983, ApJ 271, 188

\bibitem[1996]{ka96}
Kastner J.H., Weintraub D.A., Gatley I., Merrill K.M., Probst R.G. 1996,
ApJ 462, 777

\bibitem[1986]{La86}
Lambert D.L., Gustafsson B., Eriksson K., Hinkle K.H. 1986, ApJS 62, 373

\bibitem[1993]{LP93}
Langer W.D., Penzias A.A. 1993, ApJ 408, 539

\bibitem[1998]{laverny}
de Laverny, P. \& Gustafsson, B. 1998, A\&A 332, 661

\bibitem[1988]{LFOM88}
Likkel L., Forveille T., Omont A.,  Morris M. 1988, A\&A 198, L1

\bibitem[1996]{manch}
Manchado A., Guerrero M., Stanghellini L., Serra--Ricart M. 1996, 
The IAC Morphological Catalog of Northern Galactic Planetary Nebulae 
(Tenerife: IAC)

\bibitem[1989]{marigo}
Marigo P. 1998, PhD  Thesis, University of Padua

\bibitem[1989]{masso}
Masson C. R. 1989, ApJ 336, 294

\bibitem[1999]{ml99}
Mullan D. J., Linsky J. L. 1999, ApJ 511, 502

\bibitem[1996]{0na96}
Ohnaka K., Tsuji T. 1996, A\&A 310, 933

\bibitem[1998]{0na98}
Ohnaka K., Tsuji T. 1998, A\&A 335, 1018

\bibitem[1997]{pila97}
Pilachowski C.A., Sneden C., Hinkle K., Joyce R. 1997, AJ 114, 819

\bibitem[1996]{potta96}
Pottasch S. R. 1996, A\&A 307, 561 

\bibitem[1992]{RBW}
Rood R.T., Bania T.M., Wilson T.L. 1992, Nat 355, 618

\bibitem[1995]{RBWB}
Rood R.T., Bania T.M., Wilson T.L., Balser D.S. 1995, in ESO-EIPC Workshop
on the light elements, ed. P. Crane (Heidelberg: Springer), p.201

\bibitem[1999]{RBWB99}
Rood R.T., Bania T.M., Balser D.S., Wilson T.L.  1998, SSRv 84, 185

\bibitem[1998]{S98}
Sahai R., Trauger J.T., Watson A.M., et al. 1998, ApJ 493, 301

\bibitem[1998]{SVT98}
Sandrelli S., Visco A., Tosi M. 1998, in Nuclear Astrophysics, eds.
M. Arnould et al. (New York: AIP Conf. Proceedings), 425, p.581

\bibitem[1992]{Sch92}
Schwarz H. E., Corradi R.L.M., Melnik J. 1992, A\&AS 96, 23

\bibitem[1986]{Sne86}
Sneden C., Pilachowski C.A., Vandenberg D.A. 1986, ApJ 311, 826

\bibitem[1993]{SR93}
Stanghellini L., Renzini A. 1993, in IAU Symp. 155, Planetary Nebulae,
eds. R. Weinberger, A. Acker (Dordrecht: Kluwer), p.473

\bibitem[1995]{Str95}
Straniero O., Gallino R., Busso M., et al. 1995, ApJ 440, L85

\bibitem[1996]{T96}
Tosi M. 1996, in From Stars to Galaxies, eds. C. Leitherer, U. Fritze-von 
Alvensleben, J. Huchra (ASP Conf. Ser.), 98, p.299

\bibitem[1997]{VG97}
van den Hoek L.B., Groenewegen M.A.T. 1997, A\&AS 123, 305 (HG)

\bibitem[1997]{ws97}
Wannier P.G., Sahai R. 1987 ApJ 319, 367

\bibitem[1995]{wbs}
Wasserburg G.J., Boothroyd A.I., Sackmann I.-J. 1995, ApJ 447, L37

\bibitem[1987]{Wei87}
Weidemann V. 1987, A\&A 188, 74

\bibitem[1996]{WWD}
Weiss A., Wagenhuber J., Denissenkov P.A. 1996, A\&A 313, 581

\bibitem[1995]{WW95}
Woosley S.E., Weaver T.A. 1995, ApJS 101, 181

\bibitem[1996]{WB96}
Wouterloot J.G.A., Brand J. 1996, A\&AS 119, 439

\bibitem[1997]{ZNBBG97}
Zweigle J., Neri R., Bachiller R., Bujarrabal V., Grewing M. 1997, 
A\&A 324, 624

\end{thebibliography}
\end{document}